\documentclass[aps,twocolumn,floatfix]{revtex4-2}
\usepackage{amsmath,amsfonts,amssymb}
\usepackage[margin=0.7in,centering]{geometry}
\usepackage[utf8]{inputenc}
\usepackage{outlines}
\usepackage[backref=section,bookmarksopen=true]{hyperref}
\hypersetup{
	colorlinks=true,
    allcolors=[RGB]{19, 34, 114}  
}
\let\doi\undefined
\usepackage{doi}
\usepackage{bookmark}
\usepackage{graphicx}
\usepackage{blindtext}
\usepackage[normalem]{ulem}
\usepackage[all]{xy}

\newcommand{\flow}{\mu}
\newcommand{\tnd}{\tau}
\DeclareMathOperator{\prodint}{\mathcal{P}}
\newcommand{\nind}{n_{\text{ind}}}
\newcommand{\ndep}{n_{\text{dep}}}

\usepackage[caption=false]{subfig}
\usepackage{floatrow}

\usepackage{xcolor}
\usepackage{pagecolor}

\usepackage{array}
\newcolumntype{H}{>{\setbox0=\hbox\bgroup}c<{\egroup}@{}}

\allowdisplaybreaks
\begin{document}


\title{\bf Algebraic structure of hierarchic first-order reaction networks
applicable
to models of clone size distribution and stochastic gene expression
}

\author{Ximo Pechuan-Jorge$^{1}$, Raymond S. Puzio$^{4,\ast}$, Cameron Smith$^{2,3\ast}$}

\affiliation{
$^1$Department of Cancer Immunology, Genentech, Inc., 1 DNA Way, South San Francisco, CA, USA\\
$^2$Molecular Pathology Unit and Center for Cancer Research, Massachusetts General Hospital Research Institute, Department of Pathology, Harvard Medical School, Boston, MA, USA\\
$^3$ Broad Institute of MIT and Harvard, Cambridge, MA, USA\\
$^4$Hyperreal Enterprises, Ltd, 114A New Street, Musselburgh, Scotland, EH216LQ, UK\\
$\ast$ To whom correspondence should be addressed: \href{mailto:rsp@hyperreal.enterprises}{rsp@hyperreal.enterprises}, \href{mailto:cameron.smith@mgh.harvard.edu}{cameron.smith@mgh.harvard.edu}, \href{mailto:cameron@broadinstitute.org}{cameron@broadinstitute.org}. Author order is alphabetical.
}

\date{\today}

\begin{abstract}
In biology, stochastic branching processes with a two-stage, hierarchical structure arise in the study of population dynamics, gene expression, and phylogenetic inference. These models have been commonly analyzed using generating functions, the method of characteristics and various perturbative approximations. Here we describe a general method for analyzing hierarchic first-order reaction networks using Lie theory.  Crucially, we identify the fact that the Lie group associated to hierarchic reaction networks decomposes as a wreath product of the groups associated to the subnetworks of the independent and dependent types. After explaining the general method, we illustrate it on a model of population dynamics and the so-called two-state or telegraph model of single-gene transcription. Solutions to such processes provide essential input to downstream methods designed to attempt to infer parameters of these and related models.

\end{abstract}

\maketitle
\vspace{-0.8cm}
\tableofcontents

\section{Introduction}\label{sec:intro}

First-order reaction networks have been used to model biological systems at multiple levels of organization from molecular to ecological \cite{Kimmel2013-rw,Feller1968-jh,Feller1967-kr,Athreya1972-sq,Van_Kampen2007-bs,Kendall1948-jd,Nee1994-vr,Antal2010-mk, Lambert2013-ue,Roshan2014-el, Nicholson2016-xr,Durrett2010-yc, Durrett2011-bi,Durrett2013-qd,Durrett2013-cf,Durrett2015-ba,Avanzini2019-xz,Dinh2020-yt}. 
While sufficiently complex to capture interesting phenomena, they are nonetheless tractable and, in some instances, can even be solved exactly. 

Since the nineteenth century, it has been recognized that Lie theory provides a framework for systematically deriving, classifying, and understanding exact solutions to differential equations \cite{Cantwell2002-dv}.  House showed how Lie algebraic methods can be used to solve time-inhomogeneous Markov chains and illustrated his methods on several instances of first-order reaction networks \cite{House2012-hm}.  More recently, Greenman showed that an \(\mathfrak{su}(1,1)\) algebra underlies birth-death processes with quadratic rates \cite{Greenman2022-ok}. 

Proceeding in a different direction, Reis \emph{et al} noted that a salient feature of many of the solvable instances of first-order reaction networks is that they are hierarchic \cite{Reis2018-un}.  By the term ``hierarchic'', they mean that the types of individuals may be partitioned into two classes referred to as independent and dependent types. Whereas an individual whose type belongs to the former class can produce any type of offspring, an individual whose type belongs to the latter class can only produce offspring whose type also belongs to the same class.  By studying the structure of the relevant differential equations, they developed a general solution method applicable to the linear subclass of hierarchic reaction networks.

In the current investigation, we bring these two concepts together developing a general technique for analyzing hierarchic reaction networks using Lie theory.
After reviewing background material (\autoref{sec:background}), we will show how the hierarchic decomposition of types manifests itself as a decomposition of the algebraic structure of the model (\autoref{sec:decomposition}). A related decomposition has been studied in the context of finite state automata and applied to differentiable dynamical systems \cite{Krener1977-ln,Egri-Nagy2005-wu,Egri-Nagy2008-ng}.  Based on this decomposition, we show how to reduce the solution of any hierarchic first-order reaction system into the solution of its dependent and independent subsystems (\autoref{sec:sol}).  To illustrate the general techniques and demonstate their utility, we apply them to two of the simplest but nonetheless nontrivial models of this class, a birth-immigration-death model and a switching-emission-degradation model. 

In the birth-immigration-death (BID) model, there is one independent type \(A\), one dependent type \(B\), and three reactions
\footnote{The first reaction is often written as a pseudo-reaction \(\emptyset \to B\).  However, the resulting mathematical model is equivalent and here we prefer the form with a non-zero reactant since using it makes \autoref{eq:Zmaster}
homogeoneous.}
  \begin{align}
    A &\to A + B \\
    B &\to 2B \\
    B &\to \emptyset .
  \end{align}
This model was solved in the early days of the study of branching processes \cite{Kendall1949-kf,Kendall1950-yr}.  It has since been used extensively as a model of a growing population, for instance, in the mathematical modeling of cancer evolution \cite{Durrett2010-yc, Durrett2011-bi,Durrett2013-qd,Durrett2013-cf,Durrett2015-ba,Lian2018-uz}.

In the switching-emission-degradation (SED) model we have two independent types \(A_{1},\; A_{2}\), one dependent type \(B\) and four reactions
  \begin{align}
    A_{1} &\to A_{2} \\
    A_{2} &\to A_{1} \\
    A_{2} &\to A_{2} + B \\
    B &\to \emptyset .
  \end{align}
This model has been used to describe production of RNA by a gene that cycles stochastically between active and inactive expression states.  Its solution has been used to analyze gene expression data \cite{Raj2006-wv}. It can be solved in terms of confluent hypergeometric functions \cite{Peccoud1995-se,Iyer-Biswas2009-wh}.  In addition, it can be analyzed using techniques of perturbative QFT \cite{Albert2019-xq,Vastola2021-hg}.

\section{Lie theory for first-order reaction networks}\label{sec:background}

In this section, we introduce the generating function formulation of the class of hierarchic first order reaction networks together with its solution by the method of characteristics in a manner that emphasizes the underlying Lie theory \cite{Gilmore1974-aq,Cantwell2002-dv}. We define the generating function for reaction networks in \autoref{ss:genfunc} and the Lie algebra of the associated vector field that can be used to analyze the system in \autoref{ss:lie-algebra-vector-field}. For an introductory reference on using Lie groups and algebras to analyze dynamical systems we refer to the book by Cantwell \cite{Cantwell2002-dv}.

\subsection{Generating function for hierarchic first-order reaction networks}\label{ss:genfunc}

We describe the probability distribution of a reaction network in terms of its moment generating function \cite{Feller1967-kr,Feller1968-jh,Reis2018-un}, which is defined as
\begin{equation}\label{eq:genfuncdef}
  Z(\mathbf{x}, t) = \sum_{\mathbf{c} \in \mathbb{N}^{n}}
     P(\mathbf{c}, t) \prod_{i = 1}^{n} (x_{i})^{c_{i}}  .
\end{equation}
Here \(n\) denotes the number of types, the variable \(\mathbf{x} = (x_{1}, \ldots x_{n})\) takes values in \(\mathbb{R}^{n}\) and the variable \(\mathbf{c} = (c_{1}, \ldots c_{n})\) takes values in \(\mathbb{N}^{n}\), which are commonly used to represent counts of individuals or molecules of each type.  Here we use the phrase ``first-order'' to mean that each reaction has exactly one reactant.  As a consequence, the equation satisfied by the moment generating function is of the form
\begin{equation}\label{eq:Zmaster}
  \frac{\partial Z(\mathbf{x}, t)}{\partial t} = 
    \left( \sum_{j = 1}^{n} v_{j} (\mathbf{x}) \frac{\partial }{\partial x_{j}}  \right) Z(\mathbf{x}, t)
\end{equation}
where
 \begin{align}\label{eq:dyn-vecfield}
  v_{j} (\mathbf{x}) = \sum_{\mathbf{c} \in \mathbb{N}^{n}}
    k_{j \to c} \left( \prod_{k = 1}^{n} (x_{k})^{c_{k}} - x_{j} \right) .
\end{align}
Here, the \(k_{j \to c}\) denotes the rates of the reactions and $v_{j}$ encode the reaction velocities in terms of the generating function. See \cite{Reis2018-un} for derivation and \cite{Van_Kampen2007-bs,Schnoerr2017-as,Lazarescu2019-eo,Mendler2018-ed,Becker2020-uj} for further discussion of \autoref{eq:Zmaster}. We can interpret the \(v_{j}\)'s geometrically as components of a vector field \(v\) on \(\mathbb{R}^{n}\).

For the special case of the birth-immigration-death (BID) model, this vector field has the form
\begin{align}\label{eq:BIDvect}
    v^{\text{BID}}_1 (\mathbf{x}) &= k_{A \to A + B} \, x_1 ( x_2 - 1 ) \nonumber \\
    v^{\text{BID}}_2 (\mathbf{x}) &= (k_{B \to 2B} \, x_2 - k_{B \to \emptyset}) (x_2 - 1)
\end{align}
and for the switching-emission-degradation (SED) model, it has the form
\begin{align}\label{eq:SEDvect}
    v^{\text{SED}}_1 (\mathbf{x}) &= k_{A_1 \to A_2} \, (x_2 - x_1) \nonumber \\
    v^{\text{SED}}_2 (\mathbf{x}) &= k_{A_2 \to A_1} \, (x_1 - x_2) + 
        k_{A_2 \to A_2 + B} \, x_2 (x_3 - 1) \nonumber \\
    v^{\text{SED}}_3 (\mathbf{x}) &= k_{B \to \emptyset} \, (1 - x_3) .
\end{align}
We use these in \autoref{sec:sol} to analyze the BID and SED processes via the general approach to hierarchic first order reaction networks we develop here.

\subsection{Associated Lie algebra of vector fields}\label{ss:lie-algebra-vector-field}

Here we provide the general solution to \autoref{eq:Zmaster} via exponentiating the Lie algebra associated to the vector field \autoref{eq:dyn-vecfield}. Toward identifying the Lie algebra, let \(\mathfrak{X} (\mathbb{R}^{n})\) denote the set of smooth (\(C^{\infty})\) vector fields on \(\mathbb{R}^{n}\) (of which \autoref{eq:dyn-vecfield} is an instance) and let \(\mathsf{Diff} (\mathbb{R}^{n})\) denote the set of diffeomorphisms of \(\mathbb{R}^{n}\), an instance of which appears in the solution to \autoref{eq:Zmaster} given in \autoref{eq:gensol}.  Equipped with Lie bracket, \(\mathfrak{X} (\mathbb{R}^{n})\) becomes a Lie algebra with \(\mathsf{Diff} (\mathbb{R}^{n})\) as its corresponding Lie group \cite{Schmid2012-np}. The exponential map is given in terms of flows of vector fields---given a vector field \(v \in \mathfrak{X} (\mathbb{R}^{n})\) and a test function \(g \in C^{\infty} (\mathbb{R}^{n})\), we have
\begin{multline}\label{eq:expmapdiff}
  \exp \left( t \sum_i v_i (\mathbf{x}) \frac{\partial}{\partial x_i} \right)
    g(\mathbf{x}) = \\ g(\flow_{1} (\mathbf{x},t), \ldots, \flow_{n} (\mathbf{x},t))
\end{multline}
where the flow \(\flow \colon \mathbb{R}^{n} \times \mathbb{R} \to \mathbb{R}^{n}\) is defined by the initial value problem
\begin{align}
  \frac{d\flow_{i} (\mathbf{x},t)}{dt} &= v_{i} (\flow_{1} (\mathbf{x},t), \ldots, \flow_{n} (\mathbf{x},t)) \\
  \flow_{i} (\mathbf{x}, 0) &= x_{i} .
\end{align}

Taking \(v\) as the vector field defined in \autoref{eq:dyn-vecfield} (or one of the specializations in \autoref{eq:BIDvect} or \autoref{eq:SEDvect}),  we may express the solution of equation \autoref{eq:Zmaster} as 
\begin{align}\label{eq:gensol}
    Z(\mathbf{x}, t) &= \exp \left( t \sum_i v_i (\mathbf{x}) \frac{\partial}{\partial x_i} \right) Z(\mathbf{x}, 0) \nonumber \\
    &= Z(\flow_{1} (\mathbf{x},t), \ldots, \flow_{n} (\mathbf{x},t), 0)
\end{align}
While \autoref{eq:gensol} provides an abstract solution, we seek a more explicit representation that accounts for the hierarchic structure of the class of first order reaction networks we study. This is finally achieved in \autoref{eq:mastersol} of \autoref{sec:exp} after which we proceed to apply that solution to the specific examples of the BID model, whose solution appears in \autoref{eq:genfunc-BID-exact}, and SED model, whose solution appears in \autoref{eq:SEDsol}.

\section{Algebraic decomposition}\label{sec:decomposition}

We now examine how the hierarchic decomposition of the type space of a first-order reaction network manifests itself algebraically as a decomposition of the corresponding Lie group into a wreath product. In \autoref{ss:semid-sum-decomp} we decompose the Lie algebra of the hierarchic first order reaction network in terms of a semidirect sum. In \autoref{ss:wreath-prod-decomp} we find the corresponding decomposition of the Lie group via the wreath product. The result in \autoref{eq:wr-decomp} allows a formal decomposition of the class of hierarchic first order reaction networks into two subproblems.

\subsection{Semidirect sum decomposition of the Lie algebra}\label{ss:semid-sum-decomp}

We begin restating the definition of the hierarchic first-order reaction network more formally than in \autoref{sec:intro}. Set \(n = \nind + \ndep\) where \(\nind\) is the number of independent types and \(\ndep\) is the number of dependent types.  Label the types so that the independent types are numbered \(1, \ldots, \nind\) and the dependent types are numbered \(\nind + 1, \ldots, n\).  Then, the hierarchic property asserts that, when \(\nind < j \le n\), we can only have \(k_{j \to \textbf{c}} \neq 0\) if \(c_{i} = 0\) for \(1 \le i \le \nind\).  Thus, from \autoref{eq:dyn-vecfield}, we see that \(v_{j}\) can only depend upon 
\(x_{\nind + 1}, \ldots x_{n}\) when \(\nind < j \le n \).  

We now interpret this condition to qualify as a hierarchic first-order reaction network algebraically.
Let \(\mathfrak{n} \subset \mathfrak{X} (\mathbb{R}^{n})\) denote the set of vector fields \(u\) such that \(u_{i} = 0\) when \(\nind < i \le n\) given by 
\begin{align}
    \mathfrak{n} &= \big\{ u \in \mathfrak{X} (\mathbb{R}^{n}) \big|  u_{i} (\mathbf{x}) = 0  \quad \nind < i \le n \big\}.
\end{align}
Let \(\mathfrak{h} \subset \mathfrak{X} (\mathbb{R}^{n})\) denote the set of vector fields \(u\) such that \(u_{i} = 0\) when \(1 \le i \le \nind\) and \(u_j\) only depends upon \(x_{\nind + 1}, \ldots, x_{n}\) when \(\nind < j \le n\) given by
\begin{align}
    \mathfrak{h} &= \big\{ u \in \mathfrak{X} (\mathbb{R}^{n}) \big|  u_{i} (\mathbf{x}) = 0  \quad 1 \le i \le \nind \nonumber \\
    &\hskip 18 pt u_{j} (\mathbf{x}) = f_{j} (x_{\nind + 1}, \ldots, x_{n}) \quad
    \nind < j \le n \big\}.
\end{align}
In terms of this notation, the vector field associated to a hierarchic first-order reaction network lies in the direct sum of these sets of vector fields, \(\mathfrak{n} \oplus \mathfrak{h}\), naturally interpreted as vector spaces.

We can regard \(\mathfrak{h}\) as the embedding of \(\mathfrak{X} (\mathbb{R}^{\ndep})\) into \(\mathbb{R}^{n}\) corresponding to the inclusion \(\mathbb{R}^{\ndep} \subset \mathbb{R}^{n}\).  Likewise, we can conceive of an element \(u \in \mathfrak{n}\) as a family of vector fields on \(\mathbb{R}^{\nind}\) parameterized by \(\mathbb{R}^{\ndep}\), so \(\mathfrak{n} \cong \left(\mathfrak{X} (\mathbb{R}^{\nind})\right)^{\mathbb{R^{\ndep}}}\).  Geometrically, we have the trivial fiber bundle \(\mathbb{R}^{\nind} \to \mathbb{R}^{n} \to \mathbb{R}^{\ndep}\).  Then \(\mathfrak{h}\) corresponds to vector fields on the base space \(\mathbb{R}^{\ndep}\) whilst \(\mathfrak{n}\) corresponds to vector fields on the total space \(\mathbb{R}^{n}\) that are tangent to the fibers, which are of course copies of $\mathbb{R}^{\nind}$.

From these definitions, one can readily verify that both \(\mathfrak{n}\) and \(\mathfrak{h}\) are closed under Lie bracket and hence are Lie subalgebras of \(\mathfrak{X} (\mathbb{R}^{n})\).  Furthermore, if \(u \in \mathfrak{n}\) and \(v \in \mathfrak{h}\), then \([u, v] \in \mathfrak{n}\).  Thus, \(\mathfrak{n} \oplus \mathfrak{h}\) is also a Lie subalgebra of \(\mathfrak{X} (\mathbb{R}^{n})\) and \(\mathfrak{n}\) is an ideal of this subalgebra, so \(\mathfrak{n} \oplus \mathfrak{h}\) is indeed the semidirect sum of \(\mathfrak{n}\) and \(\mathfrak{h}\). This achieves the decomposition of the Lie algebra into the dependent $(\mathfrak{n})$ and independent $(\mathfrak{h})$ components, which we proceed to lift to the level of the Lie group in the following section.

\subsection{Wreath product decomposition of the Lie group}\label{ss:wreath-prod-decomp}
 
We now apply the exponential map to lift the decomposition from the Lie algebra to the Lie group. The image of the subalgebra \(\mathfrak{n}\) under the exponential map described in \autoref{eq:expmapdiff} is the subgroup \(N \subset \mathsf{Diff} (\mathbb{R}^{n})\) that consists of all invertible maps \(F \colon \mathbb{R}^{n} \to \mathbb{R}^{n}\), which can be expressed in the form
\begin{align}\label{eq:N-element}
    F_{i} (\mathbf{x}) &= f_i (x_{1} \ldots x_{n}), & 1 \le &i \le \nind \nonumber \\
    F_{j} (\mathbf{x}) &= x_{j}, & \nind < &j \le n
\end{align}
for some \(f \colon \mathbb{R}^{\nind} \times \mathbb{R}^{\ndep} \to \mathbb{R}^{\nind}\).
The image of the the subalgebra \(\mathfrak{h}\) under the exponential map is the subgroup \(H \subset \mathsf{Diff} (\mathbb{R}^{n})\) that consists of all invertible maps \(G \colon \mathbb{R}^{n} \to \mathbb{R}^{n}\), which can be expressed in the form
\begin{align}\label{eq:H-element}
    G_{i} (\mathbf{x}) &= x_{i}, & 1 \le &i \le \nind \nonumber \\
    G_{j} (\mathbf{x}) &= g_i (x_{\nind + 1} \ldots x_{n}), & \nind < &j \le n
\end{align}
for some \(g \colon \mathbb{R}^{\ndep} \to \mathbb{R}^{\ndep}\).
Conjugating the element \(F \in N\) by the element \(G \in H\), we obtain an element \(\tilde{F} = G^{-1} \circ F \circ G \in N\) which is explicitly given as
\begin{align}\label{eq:explicit-conjugation}
    \tilde{F}_{i} (\mathbf{x}) &= f_i (x_{1} \ldots x_{\nind},
    g_{\nind+1} (x_{\nind+1} \ldots x_{n})),  \nonumber \\ 
    &\hskip 24 pt \ldots, g_{n} (x_{\nind + 1} \ldots x_{n})), 
    &  \hskip -36 pt 1 \le &i \le \nind \nonumber \\
    \tilde{F}_{j} (\mathbf{x}) &= x_{j}, &  \hskip -36 pt \nind < &j \le n.
\end{align}

Either by direct verification or from the corresponding statements about \(\mathfrak{n}\) and \(\mathfrak{h}\), one can conclude that the product of subgroups \(NH\) is a group, that \(N\) is a normal subgroup of the group \(NH\), and so \(NH\) is the semidirect product of \(N \rtimes H\).

From \autoref{eq:H-element}, it is apparent that \(H\) is isomorphic to \(\mathsf{Diff} (\mathbb{R}^{\ndep})\).  In \autoref{eq:N-element}, we can regard \(f\) as specifying a family of maps from \(\mathbb{R}^{\nind}\) to \(\mathbb{R}^{\nind}\) parameterized by \(\mathbb{R}^{\ndep}\).  Thus, \(N\) is isomorphic to \(\mathsf{Diff} (\mathbb{R}^{\nind})^{\mathbb{R}^{\ndep}}\), the group of families of
diffeomorphisms of \(\mathbb{R}^{\nind}\) smoothly parameterized by \(\mathbb{R}^{\ndep}\) with pointwise composition.  Finally, \autoref{eq:explicit-conjugation} shows that the action of \(H\) on \(N\) by conjugation is isomorphic to the action of \(\mathsf{Diff} (\mathbb{R}^{\ndep})\) on \(\mathsf{Diff} (\mathbb{R}^{\nind})^{\mathbb{R}^{\ndep}}\) by reparameterizaton.  Thus, the semidirect product is isomorphic to the wreath product 
\begin{equation}\label{eq:wr-decomp}
N \rtimes H \cong
\mathsf{Diff} (\mathbb{R}^{\nind}) \wr_{\mathbb{R}^{\ndep}} \mathsf{Diff} (\mathbb{R}^{\ndep}).
\end{equation}
By analogy to the decomposition of the Lie algebra into the dependent $(\mathfrak{n})$ and independent $(\mathfrak{h})$ components from \autoref{ss:semid-sum-decomp}, \autoref{eq:wr-decomp} achieves the decomposition of the Lie group into the dependent $(N)$ and independent $(H)$ components. The difference is that the former decomposition is given in terms of the semidirect sum whereas the latter is in terms of the wreath product.

\section{Solution of hierarchic processes}\label{sec:sol}

For a particular instance of a reaction network, one may be able to further restrict attention to proper subalgebras of \(\mathfrak{X} (\mathbb{R}^{\nind})\) and \(\mathfrak{X} (\mathbb{R}^{\ndep})\).  When these subalgebras are finite-dimensional, as in our illustrative examples, we can produce closed-form solutions to these models. 

In \autoref{ss:sum-subalgebras}, building on the decomposition achieved in \autoref{ss:semid-sum-decomp}, we now combine the subalgebras to produce arbitrary instances of the vector fields that generate the dynamics of hierarchic first-order reaction networks. We apply this construction to the BID process in \autoref{ss:algBIDex} and SED process in \autoref{ss:algSEDex}.

In \autoref{sec:exp}, similarly building on the the decomposition achieved in \autoref{ss:wreath-prod-decomp}, we combine the subgroups in a manner that retains the modularity implicit in the decomposition. This allows us to eventually support application of a well-known and straightforward solution technique via matrix exponentiation. In \autoref{ss:groupBIDex} and \autoref{ss:groupSEDex}, we apply this construction to provide solutions to the BID and SED processes respectively.

\subsection{Semidirect sum of subalgebras}\label{ss:sum-subalgebras}

In this section we work at the level of Lie algebras prior to lifting via exponentiation to the associated Lie groups in \autoref{sec:exp}. In particular, we construct an algebra that maps to a subalgebra of vector fields so as to contain the vector field of \autoref{eq:dyn-vecfield} that generates the dynamics. If we are able to do this, we will be able to reduce the infinite-dimensional operator to a finite-dimensional one that can be solved directly via matrix exponentiation and integration.

Suppose that \(\mathfrak{a}\) and \(\mathfrak{b}\) are finite-dimensional Lie algebras with corresponding Lie groups \(\mathsf{A}\) and \(\mathsf{B}\).  Let \(\rho_{\mathfrak{a}} \colon \mathfrak{a} \to \mathfrak{X} (\mathbb{R}^{\nind})\) and \(\rho_{\mathfrak{b}} \colon \mathfrak{b} \to \mathfrak{X} (\mathbb{R}^{\ndep})\) be representations of these algebras by vector fields on the fiber space and the base space, respectively. We combine these to produce the Lie algebra \(\mathfrak{d}\) whose underlying vector space is \((C^{\infty} (\mathbb{R}^{\ndep}) \otimes \mathfrak{a}) \oplus \mathfrak{b}\) and whose Lie products are defined as 
\begin{align}\label{eq:d-brackets}
    [f_1 \otimes a_1, f_1 \otimes a_2]_{\mathfrak{d}} &= f_1 f_2 \otimes [a_1, a_2]_{\mathfrak{a}} \nonumber \\
    &\qquad a_1, a_2 \in \mathfrak{a}; f_1, f_2 \in C^{\infty} (\mathbb{R}^{\ndep}) \nonumber \\
    [b, f \otimes a]_{\mathfrak{d}} &= \sum_{k = 1}^{\ndep} (\rho_{\mathfrak{b}} (b))_{k} \frac{\partial f}{\partial x_{k}} \otimes a \nonumber \\ 
    &\qquad b \in \mathfrak{b}; a \in \mathfrak{a}; f \in C^{\infty} (\mathbb{R}^{\ndep}) \nonumber \\
    [b_1, b_2]_\mathfrak{\mathfrak{d}} &= [b_1, b_2]_{\mathfrak{b}} \\
    &\qquad b_1, b_2 \in \mathfrak{b} \nonumber .
\end{align}
From the above, we see that \(C^{\infty} (\mathbb{R}^{\ndep}) \otimes \mathfrak{a}\) is an ideal and that \(\mathfrak{d}\) is a semidirect sum of this ideal with \(\mathfrak{b}\).

Note that since \(\mathfrak{a}\) is finite-dimensional, we have \(C^{\infty} (\mathbb{R}^{\ndep}) \otimes \mathfrak{a} \cong \mathfrak{a}^{\mathbb{R}^{\ndep}}\).  Hence, 
the construction of \(\mathfrak{d}\) can be understood as combining the algebras \(\mathfrak{a}\) and \(\mathfrak{b}\) to produce a semidirect sum \(\mathfrak{a}^{\mathbb{R}^{\ndep}} \oplus \mathfrak{b}\) analogous to the manner in which \(\mathfrak{X} (\mathbb{R}^{\nind})\) and \(\mathfrak{X} (\mathbb{R}^{\ndep})\) combine to form the algebra \(\mathfrak{n} \oplus \mathfrak{h} \cong \mathfrak{X} (\mathbb{R}^{\nind})^{\mathbb{R}^{\ndep}} \oplus \mathfrak{X} (\mathbb{R}^{\ndep})\).

To make this analogy explicit, we define a representation \(\rho_{\mathfrak{d}} \colon \mathfrak{d} \to \mathfrak{n} \oplus \mathfrak{h}\) of this combined algebra on vector fields.  To each element \(f \otimes a \in C^{\infty} (\mathbb{R}^{\ndep}) \otimes \mathfrak{a}\), we associate the vector field \(u \in \mathfrak{n}\) defined as
\begin{align}\label{eq:rhoA}
    u_{i} (x_{1}, \ldots x_{n}) &= (\rho_{\mathfrak{a}} (a))_{i} (x_{1}, \ldots , x_{\nind})
        f (x_{\nind + 1}, \ldots, x_{n}), \nonumber \\
        &\qquad 1 \le i \le \nind \nonumber \\
    u_{i} (x_{1}, \ldots , x_{n}) &= 0, \nonumber \\
        &\qquad  \nind < i \le n 
\end{align}
and to each element \(b \in \mathfrak{b}\) we associate the vector field \(w \in \mathfrak{n}\) defined as
\begin{align}\label{eq:rhoB}
    w_{i} (x_{1}, \ldots x_{n}) &= 0, \nonumber \\
        &\qquad 1 \le i \le \nind \nonumber \\
    w_{i} (x_{1}, \ldots x_{n}) &= (\rho_{\mathfrak{b}} (b))_{i} (x_{\nind + 1}, \ldots x_{n}), \nonumber \\
        &\qquad  \nind < i \le n. 
\end{align}
We will now see how this construction allows us to recover the vector fields associated to particular instances of reaction networks.

\subsubsection{Example of BID process}\label{ss:algBIDex}

For this example, we will take \(\mathfrak{a}\) to be the Lie algebra of \(1 \times 1\) matrices and we take \(\mathfrak{b}\) to be the Lie algebra of traceless \(2 \times 2\) matrices.  (\emph{i.e.\ } \(\mathfrak{a} \cong \mathfrak{gl} (1, \mathbb{R}) \cong \mathbb{R}\) and \(\mathfrak{b} \cong \mathfrak{sl} (2, \mathbb{R})\))  We represent them as vector fields on the real line,
\begin{align}
    \left(\rho_{\mathfrak{a}} (a) \right)_{1} (x) &= a_{00} x \\
    \left(\rho_{\mathfrak{b}} (M) \right)_{1} (x)
         &= M_{10} x^2 + 2 M_{00} x - M_{01} .
\end{align}
In this case, \(C^{\infty} (\mathbb{R}) \otimes \mathfrak{a} \cong C^{\infty} (\mathbb{R})\), so an element of \(\mathfrak{d}\) consists of a pair \((f,M)\) where \(f \in C^{\infty} (\mathbb{R})\) is a function and \(M \in \mathfrak{b}\) is a matrix.  Then \autoref{eq:d-brackets} becomes
\begin{align}
    [(f_1, 0), (f_2, 0)]_{\mathfrak{d}} &= (0, 0) \nonumber \\
    [(0, M), (f, 0)]_{\mathfrak{d}} &= \bigg(\left( M_{10} x^2 + 2 M_{00} x - M_{01} \right) 
        \frac{\partial f}{\partial x}, 0 \bigg) \nonumber \\
    [(0, M_1), (0, M_2)]_{\mathfrak{d}} &= (0, [M_1, M_2]_{\mathfrak{b}}) \nonumber \\
        &= (0, M_1 M_2 - M_2 M_1)
\end{align}
The representation \(\rho_{\mathfrak{d}}\) from \autoref{eq:rhoA} and \autoref{eq:rhoB} works out to be
\begin{align}
    \left( \rho_{\mathfrak{d}} (f, M) \right)_{1} (x_1, x_2) &= x_1 f(x_2) \nonumber \\ 
    \left( \rho_{\mathfrak{d}} (f, M) \right)_{2} (x_1, x_2) &= 
        M_{10} x_2^2 + 2 M_{00} x_2 - M_{01} .
\end{align}
For the particular choices
\begin{align}
    f^{\mathrm{BID}} (x) &= k_{A \to A + B}  (x - 1) \\
    M^{\mathrm{BID}} &= \begin{pmatrix}
        - \tfrac{1}{2} k_{B \to 2B} - \tfrac{1}{2} k_{B \to \emptyset} & k_{B \to \emptyset} \\
        k_{B \to 2B} & \tfrac{1}{2} k_{B \to 2B} + \tfrac{1}{2} k_{B \to \emptyset}
    \end{pmatrix},
\end{align}
this reproduces the vector field \(v^{\text{BID}} = \rho_{\mathfrak{d}} (f^{\mathrm{BID}}, M^{\mathrm{BID}})\) in \autoref{eq:BIDvect}.

\subsubsection{Example of SED process}\label{ss:algSEDex}

For this example, we will take \(\mathfrak{a}\) to be the Lie algebra of \(2 \times 2\) matrices and we take \(\mathfrak{b}\) to be the Lie algebra of traceless triangular \(2 \times 2\) matrices.  (\emph{i.e.\ } \(\mathfrak{a} \cong \mathfrak{gl} (2, \mathbb{R}) \cong \mathbb{R}\) and \(\mathfrak{b} \cong \mathfrak{igl} (1, \mathbb{R})\)).

We represent them as vector fields on \(\mathbb{R}^{2}\) and \(\mathbb{R}\) respctively,
\begin{align}
    \left(\rho_{\mathfrak{a}} (M) \right)_{1} (x_1, x_2, x_3) &= M_{11} x_{1} + M_{21} x_{2} \nonumber \\
    \left(\rho_{\mathfrak{a}} (M) \right)_{2} (x_1, x_2, x_3) &= M_{12} x_{1} + M_{22} x_{2} \nonumber \\
    \left(\rho_{\mathfrak{b}} (M) \right)_{1} (x_1, x_2, x_3) &= 2 M_{11} x_3 - M_{12} 
\end{align}
In this case, \(C^{\infty} (\mathbb{R}) \otimes \mathfrak{a}\) consists of \(2 \times 2\) matrices of real functions, so an element of \(\mathfrak{d}\) consists of a pair \((A, B)\) where \(A\) is a matrix of functions and \(B\) is a matrix of numbers.   The Lie products are given as
\begin{align}
    [(A,0), (B,0)]_{\mathfrak{d}} &= (AB - BA, 0) \nonumber \\
    [(0,A), (B,0)]_{\mathfrak{d}} &= \left( (2 B_{11} x_3 - B_{12}) \frac{\partial A}{\partial x_{3}}, 0 \right) \nonumber \\
    [(0, A), (0, B)] &= (0, AB - BA) .
\end{align}
The represenatation \(\rho_{\mathfrak{d}}\) becomes
\begin{align}
    \left( \rho_{\mathfrak{d}} (A, B) \right)_{1} (x_1, x_2, x_3) &= A_{11} (x_3) x_{1} + A_{21} (x_3) x_{2} \nonumber \\
    \left( \rho_{\mathfrak{d}} (A, B) \right)_{2} (x_1, x_2, x_3) &= A_{12} (x_3) x_{1} + A_{22} (x_3) x_2 \nonumber \\
    \left( \rho_{\mathfrak{d}} (A, B) \right)_{3} (x_1, x_2, x_3) &= 2 B_{11} x_3 - B_{12} .
\end{align}
For the choice
\begin{align}
    A^{\text{SED}} &= \begin{pmatrix}
           -k_{A_1 \to A_2} & k_{A_1 \to A_2} \\
            k_{A_2 \to A_1} & k_{A_2 \to A_2 + B} (x_3 - 1) - k_{A_2 \to A_1} 
         \end{pmatrix} \label{eq:ASED} \\
    B^{\text{SED}} &= k_{B \to 0} 
        \begin{pmatrix}
            \tfrac{1}{2} & 1 \\ 
            0 & -\tfrac{1}{2}
        \end{pmatrix} \label{eq:BSED}
\end{align}
we recover the vector field \(v^{\text{SED}} = \rho_{\mathfrak{d}} (A^{\text{SED}}, B^{\text{SED}})\) of \autoref{eq:SEDvect}.

\subsection{Semidirect product of subgroups}\label{sec:exp}

Denote the group representations to which \(\rho_{\mathfrak{a}}\) and \(\rho_{\mathfrak{b}}\) exponentiate as \(\phi_{\mathfrak{a}} \colon \mathsf{A} \to \mathsf{Diff} (\mathbb{R}^{\nind})\) and \(\phi_{\mathfrak{b}} \colon \mathsf{B} \to \mathsf{Diff} (\mathbb{R}^{\ndep})\).   

We construct the group \(\mathsf{D}\) corresponding to the Lie algebra \(\mathfrak{d}\) described above as a wreath product \(\mathsf{D} = \mathsf{A}^{\mathbb{R}^{\ndep}} \wr_{\phi_{\mathfrak{b}}} \mathsf{B}\).  Let \(\mathsf{A}^{\mathbb{R}^{\ndep}}\) denote the set of smooth maps from \(\mathbb{R}^{\ndep}\) to \(\mathsf{A}\), which forms a group under pointwise multiplication.  The underlying set of \(\mathsf{D}\) is  
\(\mathsf{A}^{\mathbb{R}^{\ndep}} \times \mathsf{B}\) and multiplication is given as
\begin{equation}
    (\alpha_1, \beta_1) \cdot (\alpha_2, \beta_2) = 
    (\alpha_1 \cdot (\alpha_2 \circ \rho_{\mathfrak{b}} (\beta_1), \beta_1 \cdot \beta_2)) 
\end{equation}
for \(\alpha_1, \alpha_2 \in \mathsf{A}^{\mathbb{R}^{\ndep}}\) and \(\beta_1, \beta_2 \in \mathsf{B}\).

Next, we construct the exponential map from \(\mathfrak{d}\) to \(\mathsf{D}\).  For \(b \in \mathfrak{b}\), we have \(\exp(0, b) = (\mathbf{id}, \exp(x)) \in \mathsf{D}\).  For \(a \in \mathfrak{a}^{\mathbb{R}^{\ndep}}\), we exponentiate pointwise, so
\begin{align}
    \exp(a) (p) &= \exp(a(p)) \nonumber \\
    \exp(a, 0) &= (\exp (a), \mathbf{id}) \in \mathsf{D} .
\end{align}
To combine these, we will make use of the fundamental identity
\begin{multline}\label{eq:prodintpart}
  \exp (u + v) =  \\
  \exp (x) \prodint_{0}^{1} \exp \left\{ dt \, \exp (-tu) \, v \, \exp (tu) \right\} .
\end{multline}
Here \(\prodint\) indicates the product integral and \(x,y\) are elements of a Lie algebra (sec.\ A.II.10 of \cite{Masani1984-ql,Dollard1984-to}).  From the defining equation, \autoref{eq:d-brackets}, we see that
\begin{align}
    &\exp ([b,-]) (f \otimes a) \nonumber \\
    &\hskip 18pt = \left( \exp \left\{ \sum_{k = 1}^{\ndep} (\rho_{\mathfrak{b}} (b))_{k} \frac{\partial }{\partial x_{k}} \right\} f \right) \otimes v \nonumber \\
    &\hskip 18pt = (f \circ \phi_{\mathfrak{b}} (b)) \otimes a
\end{align}
Hence, by the identity
\begin{equation}
    \exp (-tu) \, v \, \exp (tu) = \exp ([-,u]) v,
\end{equation}
we have \(\exp (-tb) \, a \, \exp (tb) = a \circ \phi_{\mathfrak{b}} (tb)\), so conjugation by \(\exp(y)\) corresponds to acting on \(\mathbb{R}^{\ndep}\).  Combining these observations, we conclude that
\begin{multline}
  \exp ((a, b)) = \\ \bigg(
  \prodint_{0}^{1} \exp \left\{dt \,  a \circ \phi_{\mathfrak{b}} \big( \exp (tb) \big) \right\},
  \exp(b) \bigg) .
\end{multline}

Corresponding to \(\rho_{\mathfrak{d}}\), there is an action \(\phi_{\mathfrak{d}} \colon \mathsf{D} \to (\mathbb{R}^{\nind}) \wr_{\mathbb{R}^{\ndep}} \mathsf{Diff} (\mathbb{R}^{\ndep})\).  Given \((\alpha, \beta) \in \mathsf{D}\), we have
\begin{multline}
    (\phi_{\mathfrak{d}} (\alpha, \beta)) (x_{1}\ldots, x_{n}) = \\ 
        \big( \phi_{\mathfrak{a}} (\alpha (x_{\nind + 1}, \ldots, x_{n})) (x_{1}\ldots, x_{\nind}), \\
        \phi_{\mathfrak{b}} (\beta) (x_{\nind + 1}, \ldots, x_{n}) \big)
\end{multline}

Using this action, we can express the exponential of the vector field corresponding to an element \((a,b) \in \mathfrak{d}\) as
\begin{equation}
    \exp \left(\sum_{k = 1}^{n} \big(\rho_{\mathfrak{d}} ((a,b))\big)_{k} 
        \frac{\partial}{\partial x_{k}} \right) =
    \phi_{\mathfrak{d}} \big(\exp((a,b))\big)
\end{equation}

In particular, if \((a,b)\) is the element corresponding to the vector field which generates the dynamics (such as \((f^{\text{BID}}, M^{\text{BID}})\) or \((A^{\text{SED}}, B^{\text{SED}})\) above) we can use this expression to solve for the generating function.
Substituting into in \autoref{eq:gensol}, we obtain
\begin{align}\label{eq:mastersol}
    Z(\mathbf{x}, t) &= Z(\phi_{\mathfrak{d}} (\exp ((a, b)t)) (\mathbf{x}), 0) \nonumber \\ 
    &= Z\bigg(
    \phi_{\mathfrak{d}} \left( \prodint_{0}^{t} \exp \left\{dt' \,  a \circ \phi_{\mathfrak{b}} \big( \exp (t'b) \big) \right\} \right) \\
    &\hskip 60 pt \left( \phi_{\mathfrak{d}} (\exp (tb) (\mathbf{x})) \right), 0 \bigg) \nonumber
\end{align}
as the solution to \autoref{eq:Zmaster}.

\subsubsection{Example of BID process}\label{ss:groupBIDex}

For the BID model, the Lie group \(\mathsf{A} \cong \mathsf{GL}(1, \mathbb{R})\) corresponding to \(\mathfrak{a}\) consists of invertible \(1 \times 1\) matrices and the Lie group \(\mathsf{B} \cong \mathsf{SL}(1, \mathbb{R})\) corresponding to \(\mathfrak{b}\) consists of \(2 \times 2\) matrices with unit determinant.  These groups are represented by diffeomorphisms acting on \(\mathbb{R}^{\nind}\) and on \(\mathbb{R}^{\ndep}\) as
\begin{align}\label{eq:phiBID}
    \phi_{\mathfrak{a}} (A) (x) &= A x \\
    \phi_{\mathfrak{b}} (B) (x) &= \frac{B_{00} x + B_{01}}
                                        {B_{10} x + B_{11}} .
\end{align}
These actions consist of dilations of the real line and fractional linear transorms, respectively.

For the particular choices \(f^{\mathrm{BID}}\) and \(M^{\mathrm{BID}}\), we obtain 
\begin{multline}
    \phi_{\mathfrak{b}} (\exp (t M^{\mathrm{BID}})) (x) = \\
    \frac{\begin{aligned} (1 &- \sigma) \, (1 - e^{-\tnd}) \\[-4pt] &+ 
            ((1 - \sigma) \, e^{-\tnd} - 1 - \sigma) x \end{aligned}}
        {\begin{aligned} 1 - \sigma &- (1 + \sigma) \, e^{-\tnd} \\[-4pt] &-
        (1 + \sigma) \, (1 - e^{-\tnd}) x \end{aligned}} .
\end{multline}
where, for convenience, we define the quantities
\begin{align}
    \sigma &= \frac{k_{B \to 2B} - k_{B \to \emptyset}}
        {k_{B \to 2B} + k_{B \to \emptyset}} \\
    m &= \frac{k_{A \to A + B}}{k_{B \to 2B} - k_{B \to \emptyset}} \\
    \tnd &= (k_{B \to 2B} - k_{B \to \emptyset}) t
\end{align}
Since \(\mathfrak{a}\) is commutative, the time-ordered integral reduces to an ordinary integral and we have
\begin{align}
     &\prodint_{0}^{t} \exp \left\{dt' \, f^{\mathrm{BID}} \circ \phi_{\mathfrak{b}} \big( \exp (t' M^{\mathrm{BID}}) (x_2) \big) \right\} \nonumber \\
     &\hskip 24pt = \exp \left\{\int_{0}^{t} dt' \, f^{\mathrm{BID}} \circ \phi_{\mathfrak{b}} \big( \exp (t' M^{\mathrm{BID}}) (x_2) \big) \right\} \nonumber \\
      &\hskip 24pt = \left( \frac{2 \sigma e^{-\tnd}}
        {\begin{aligned} &(1 + \sigma)(1 - e^{-\tnd}) \, x_2 \\[-4pt] &\quad 
          - 1 - \sigma + (1 - \sigma) e^{-\tnd} \end{aligned}}
    \right)^{\frac{2 m \sigma}{1 + \sigma}}
\end{align}
For the initial condition \(Z(x_1, x_2, 0) = x_{1} x_{2}^{n_0}\), we obtain the solution
\begin{align}\label{eq:genfunc-BID-exact}
    Z (x_1, x_2, \tnd) &= x_1 \,
    \left( \frac{\begin{aligned} (\sigma &- 1)(1 - e^{-\tnd}) \\[-4pt]
   &+ (1 -\sigma - (1 +\sigma) e^{-\tnd}) x \end{aligned}}
        {\begin{aligned} -1 &- \sigma + (1 - \sigma) e^{-\tnd} \\[-4pt] &+
        (1 + \sigma) (1 - e^{-\tnd}) \, x_2 \end{aligned}} \right)^{n_{0}} \nonumber \\ &\qquad\times
    \left( \frac{2 \sigma e^{-\tnd}}
        {\begin{aligned} &(1 + \sigma)(1 - e^{-\tnd}) \, x_2  \\[-4pt] &\quad 
          - 1 - \sigma + (1 - \sigma) e^{-\tnd} \end{aligned}}
    \right)^{\frac{2 m \sigma}{1 + \sigma}}
\end{align}
from the general solution \autoref{eq:mastersol}.  This reproduces the well-known solution of this model.

\subsubsection{Example of SED process}\label{ss:groupSEDex}

For the SED model, the Lie group \(\mathsf{A} \cong \mathsf{GL}(2, \mathbb{R})\) corresponding to \(\mathfrak{a}\) consists of invertible \(2 \times 2\) matrices and the Lie group \(\mathsf{B} \cong \mathsf{IGL}(1, \mathbb{R})\) corresponding to \(\mathfrak{b}\) consists of \(2 \times 2\) triangular matrices with unit determinant.  Elements of these groups \(A \in \mathsf{A}\) and \(B \in \mathsf{B}\) are represented by diffeomorphisms acting on \(\mathbb{R}^{\nind}\) and on \(\mathbb{R}^{\ndep}\) as
\begin{align}
    \phi_{\mathfrak{a}} (A) (x_1, x_2) &= A \begin{pmatrix} x_1 \\ x_2 \end{pmatrix} \label{eq:phiASED} \\
    \phi_{\mathfrak{b}} (B) (x_3) &= B_{00} x_3 + B_{01}. \label{eq:phiBSED}
\end{align}
These actions consist of linear transforms of the plane and affine transforms transforms of the real line, respectively.

Computing the exponential of \autoref{eq:BSED},
\begin{equation}
    \exp(t B^{\text{SED}}) =
    \begin{pmatrix}
        e^{\tfrac{1}{2} k_{B \to \emptyset} t} & 
            2(e^{-\tfrac{1}{2} k_{B \to \emptyset} t} - 1) \\
            0 & e^{-\tfrac{1}{2} k_{B \to \emptyset} t}
    \end{pmatrix} ,
\end{equation}
so, by \autoref{eq:phiBSED}, we have
    \begin{equation}
    \phi_{\mathfrak{b}} (\exp(t B^{\text{SED}})) (x) =
        e^{\tfrac{1}{2} k_{B \to \emptyset} t} x +
        2(e^{-\tfrac{1}{2} k_{B \to \emptyset} t} - 1) .
\end{equation}
Substituting this into \autoref{eq:ASED} produces
\begin{equation}
    A^{\text{SED}} \circ \phi_{\mathfrak{b}} (\exp(t' B^{\text{SED}})) (x_3) =
    M_{1} \xi + M_{2}
\end{equation}
where
\begin{align}
    M_{1} &= \begin{pmatrix}
           -k_{A_1 \to A_2} & k_{A_1 \to A_2} \\
            k_{A_2 \to A_1} & - k_{A_2 \to A_1} - k_{A_2 \to A_2 + B}
         \end{pmatrix} \nonumber \\
    M_{2} &= \begin{pmatrix}
        0 & 0 \\ 0 & k_{A_2 \to A_2 + B}
    \end{pmatrix} \nonumber \\
    \xi &= e^{\tfrac{1}{2} k_{B \to \emptyset} t'} x_3 +
        2(e^{-\tfrac{1}{2} k_{B \to \emptyset} t'} - 1).
\end{align}
Taking the initial condition, 
\begin{equation}
    Z(x_1, x_2, x_3, 0) = 
    (p x_1 + (1 - p) x_2) x_3^{n_0},
\end{equation}
we obtain the solution 
\begin{align}\label{eq:SEDsol}
    &Z(x_1, x_2, x_3, t) = \nonumber \\
    &\quad\left( e^{\tfrac{1}{2} k_{B \to \emptyset} t} x_3 +
        2(e^{-\tfrac{1}{2} k_{B \to \emptyset} t} - 1) \right)^{n_0} \times \nonumber \\
    &\quad\begin{pmatrix}
        p & 1-p
    \end{pmatrix}
    \prodint_{0}^{t}
    \exp\left\{ (M_1 \xi + M_2) \, dt' \right\}
    \begin{pmatrix}
        x_1 \\ x_2
    \end{pmatrix} .
\end{align}

Finally, we indicate how this form of the solution in \autoref{eq:SEDsol} relates to the form of the solution found in \cite{Peccoud1995-se,Iyer-Biswas2009-wh}.
Making a change of variable \(z = e^{\tfrac{1}{2} k_{B \to \emptyset} t}\), we obtain
\begin{multline}
    \prodint_{0}^{t} \exp \left\{ A^{\text{SED}} \circ \phi_{\mathfrak{b}} \big(\exp(t' B^{\text{SED}}) \big) \,  dt' \right\} \\
     = \prodint_{0}^{z} \exp \left\{\frac{2}{k_{B \to \emptyset} z'} 
        (M_1 z' + M_2) \, dz' \right\} = X .
\end{multline}
We note that this quantity is the solution to the matrix differential equation
\begin{equation}
    \frac{d X}{dz} = \frac{2}{k_{B \to \emptyset} z} (M_1 z + M_2) X .
\end{equation}
As explained in Chapter 4 of \cite{Hochstadt2012-xh}, this equation has a regular singularity at (z = 0) and is analytic for all \(z \in \mathbb{C} \setminus \{0\}\).  Making the change of variable \(w = 1/z\), this equation becomes
\begin{equation}
    \frac{d X}{dw} = -\frac{2}{k_{B \to \emptyset} w} \left(M_1 \frac{1}{w} + M_2 \right) X .
\end{equation}
In the terminology of Ince (\cite{Ince1956-yd}, section 20.32), this equation has an irregular singularity of the second species at \(z = \infty\).  Therefore, this equation has the signature of singularities \([0, 1, 1_2]\).  As a linear differential equation with analytic coefficients, it is uniquely determined by the nature of its singularities.  In this case, the signature indicates that it is equivalent to the confluent hypergeometric equation. This is why confluent hypergeometric functions provide solutions to the generating function of the SED process.  Indeed, substituting the appropriate expression in terms of confluent hypergeometric functions for the product integral in \autoref{eq:SEDsol} recovers the solution of \cite{Iyer-Biswas2009-wh}.

\section{Discussion}

We have shown how the dynamic Lie group of a hierarchical first-order reaction network decomposes as a wreath product of groups associated to the independent and dependent subsystems, leading to an expression for the moment generating function in terms of a multiplicative integral.  This extends the scope of Lie-theoretic methods for solving reaction networks---whereas prior work required a finite-dimensional Lie group \cite{House2012-hm,Greenman2021-qh}, we employ infinite-dimensional groups built up from finite-dimensional components. This result has practical and theoretical consequences.

Practically, it can be used to obtain explicit solutions, as we demonstrated by rederiving the solutions to the BID and SED models.  Furthermore, as noted in \cite{House2012-hm}, Lie theoretic techniques and matrix exponentiation can be implemented numerically leading to efficient algorithms \cite{Keeling2008-hw,Al-Mohy2010-nf}. Ultimately, where analytic solutions are available, numerical implementations would ideally support direct comparison to approximations \cite{Ham2020-ij}. Of course, in the event that efficient sampling algorithms can be developed solutions of this sort enable probabilistic parameter inference \cite{Schnoerr2017-as,Ocal2019-ht,Dinh2020-yt,Watson2020-fc,Sukys2022-fg,Carilli2023-ti}. Hopefully, our results support contextualization if not expansion of the domain of applicability of the analytic methods we have exemplified here.

Theoretically, this Lie-algebraic account of hierarchic first-order reaction networks serves as a unifying principle. Indeed, it allowed us to systematically derive a general solution of the hierarchic first-order system which Ries \emph{et. al} \cite{Reis2018-un} first obtained by direct manipulation of the differential equations.  They also noted that the generalization to networks such as our BID example, which are hierarchic but not linear, involves the Riccati equations.  This can also be explained group theoretically---the action of \(\mathsf{PSL}(2, \mathbb{R})\) by fractional linear transforms that we encountered in \autoref{eq:phiBID} also underlies the Riccati equation \cite{Cantwell2002-dv,nlab-Riccati-2023}.  As we noted at the end of \autoref{ss:groupSEDex}, the functional form of the solution in terms of confluent hypergeometric functions follows from the structure of \autoref{eq:mastersol}.  The argument we presented generalizes to show that more complicated hierarchic networks can be solved using Fuchsian functions, as Dattani \cite{Dattani2015-kz, Dattani2017-uk} showed from the explicit differential equation.  The identity \autoref{eq:prodintpart} also serves as the basis for the interaction picture and time-dependant perturbation theory in quantum mechanics \cite{Arley1944-cr,Ticciati1999-jz}, thereby connecting to the techniques used by Vastola \emph{et. al} \cite{Vastola2021-hg}.


\bibliographystyle{unsrtnat}
\bibliography{paperpile}

\begin{thebibliography}{58}
\providecommand{\natexlab}[1]{#1}
\providecommand{\url}[1]{\texttt{#1}}
\expandafter\ifx\csname urlstyle\endcsname\relax
  \providecommand{\doi}[1]{doi: #1}\else
  \providecommand{\doi}{doi: \begingroup \urlstyle{rm}\Url}\fi

\bibitem[Kimmel and Axelrod(2013)]{Kimmel2013-rw}
Marek Kimmel and David~E Axelrod.
\newblock \emph{Branching Processes in Biology}, volume~1 of
  \emph{Interdisciplinary Applied Mathematics}.
\newblock Springer, New York, NY, October 2013.
\newblock ISBN 9781441929587.
\newblock \doi{10.1007/b97371}.

\bibitem[Feller and Feller(1968)]{Feller1968-jh}
Vilim Feller and William Feller.
\newblock \emph{An Introduction to Probability Theory and Its Applications,
  Volume 1}.
\newblock Wiley, 3rd edition, January 1968.
\newblock ISBN 9780471257080.

\bibitem[Feller(1967)]{Feller1967-kr}
William Feller.
\newblock \emph{An Introduction to Probability Theory and Its Applications,
  Volume 2}.
\newblock Wiley, 1967.
\newblock ISBN 9780471257097.

\bibitem[Athreya and Ney(1972)]{Athreya1972-sq}
Krishna~B Athreya and Peter~E Ney.
\newblock \emph{Branching Processes}, volume~1 of \emph{Grundlehren der
  Mathematischen Wissenschaften}.
\newblock Springer Berlin Heidelberg, Berlin, Germany, November 1972.
\newblock ISBN 9783540057901.

\bibitem[Van~Kampen(2007)]{Van_Kampen2007-bs}
N~G Van~Kampen.
\newblock \emph{Stochastic processes in physics and chemistry}, volume~1 of
  \emph{North-Holland Personal Library}.
\newblock North-Holland, Oxford, England, 3 edition, March 2007.
\newblock ISBN 9780444529657.
\newblock \doi{10.1016/b978-0-444-52965-7.x5000-4}.

\bibitem[Kendall(1948)]{Kendall1948-jd}
D~G Kendall.
\newblock On some modes of population growth leading to r. a. fisher's
  logarithmic series distribution.
\newblock \emph{Biometrika}, 35\penalty0 (Pts 1-2):\penalty0 6--15, May 1948.
\newblock ISSN 0006-3444.
\newblock \doi{10.2307/2332624}.

\bibitem[Nee et~al.(1994)Nee, May, and Harvey]{Nee1994-vr}
S~Nee, R~M May, and P~H Harvey.
\newblock The reconstructed evolutionary process.
\newblock \emph{Philos. Trans. R. Soc. Lond. B Biol. Sci.}, 344\penalty0
  (1309):\penalty0 305--311, May 1994.
\newblock ISSN 0962-8436.
\newblock \doi{10.1098/rstb.1994.0068}.

\bibitem[Antal and Krapivsky(2010)]{Antal2010-mk}
Tibor Antal and P~L Krapivsky.
\newblock {Exact solution of a two-type branching process: clone size
  distribution in cell division kinetics}.
\newblock \emph{J. Stat. Mech: Theory Exp.}, 2010\penalty0 (07):\penalty0
  P07028, July 2010.
\newblock ISSN 1742-5468.
\newblock \doi{10.1088/1742-5468/2010/07/P07028}.

\bibitem[Lambert and Stadler(2013)]{Lambert2013-ue}
Amaury Lambert and Tanja Stadler.
\newblock Birth-death models and coalescent point processes: the shape and
  probability of reconstructed phylogenies.
\newblock \emph{Theor. Popul. Biol.}, 90:\penalty0 113--128, December 2013.
\newblock ISSN 0040-5809, 1096-0325.
\newblock \doi{10.1016/j.tpb.2013.10.002}.

\bibitem[Roshan et~al.(2014)Roshan, Jones, and Greenman]{Roshan2014-el}
A~Roshan, P~H Jones, and C~D Greenman.
\newblock {Exact, time-independent estimation of clone size distributions in
  normal and mutated cells}.
\newblock \emph{J. R. Soc. Interface}, 11\penalty0 (99), 2014.
\newblock ISSN 1742-5689, 1742-5662.
\newblock \doi{10.1098/rsif.2014.0654}.

\bibitem[Nicholson and Antal(2016)]{Nicholson2016-xr}
Michael~D Nicholson and Tibor Antal.
\newblock {Universal Asymptotic Clone Size Distribution for General Population
  Growth}.
\newblock \emph{Bull. Math. Biol.}, 78\penalty0 (11):\penalty0 2243--2276,
  November 2016.
\newblock ISSN 0092-8240.
\newblock \doi{10.1007/s11538-016-0221-x}.

\bibitem[Durrett and Moseley(2010)]{Durrett2010-yc}
Richard Durrett and Stephen Moseley.
\newblock Evolution of resistance and progression to disease during clonal
  expansion of cancer.
\newblock \emph{Theor. Popul. Biol.}, 77\penalty0 (1):\penalty0 42--48,
  February 2010.
\newblock ISSN 0040-5809.
\newblock \doi{10.1016/j.tpb.2009.10.008}.

\bibitem[Durrett et~al.(2011)Durrett, Foo, Leder, Mayberry, and
  Michor]{Durrett2011-bi}
Rick Durrett, Jasmine Foo, Kevin Leder, John Mayberry, and Franziska Michor.
\newblock {Intratumor Heterogeneity in Evolutionary Models of Tumor
  Progression}.
\newblock \emph{Genetics}, 188\penalty0 (2):\penalty0 461--477, June 2011.
\newblock ISSN 0016-6731.
\newblock \doi{10.1534/genetics.110.125724}.

\bibitem[Durrett(2013{\natexlab{a}})]{Durrett2013-qd}
Rick Durrett.
\newblock {Population genetics of neutral mutations in exponentially growing
  cancer cell populations}.
\newblock \emph{Ann. Appl. Probab.}, 23\penalty0 (1):\penalty0 230--250,
  February 2013{\natexlab{a}}.
\newblock ISSN 1050-5164.
\newblock \doi{10.1214/11-AAP824}.

\bibitem[Durrett(2013{\natexlab{b}})]{Durrett2013-cf}
Rick Durrett.
\newblock {Cancer Modeling: A Personal Perspective}.
\newblock \emph{Not. Am. Math. Soc.}, 60\penalty0 (03):\penalty0 304, March
  2013{\natexlab{b}}.
\newblock ISSN 0002-9920.
\newblock \doi{10.1090/noti953}.

\bibitem[Durrett(2015)]{Durrett2015-ba}
Richard Durrett.
\newblock \emph{{Branching Process Models of Cancer}}.
\newblock Springer International Publishing, July 2015.
\newblock ISBN 9783319160641.
\newblock \doi{10.1007/978-3-319-16065-8\_1}.

\bibitem[Avanzini and Antal(2019)]{Avanzini2019-xz}
Stefano Avanzini and Tibor Antal.
\newblock {Cancer recurrence times from a branching process model}.
\newblock \emph{PLoS Comput. Biol.}, 15\penalty0 (11):\penalty0 e1007423,
  November 2019.
\newblock ISSN 1553-734X, 1553-7358.
\newblock \doi{10.1371/journal.pcbi.1007423}.

\bibitem[Dinh et~al.(2020)Dinh, Jaksik, Kimmel, Lambert, and
  Tavar{\'e}]{Dinh2020-yt}
Khanh~N Dinh, Roman Jaksik, Marek Kimmel, Amaury Lambert, and Simon Tavar{\'e}.
\newblock {Statistical Inference for the Evolutionary History of Cancer
  Genomes}.
\newblock \emph{Stat. Sci.}, 35\penalty0 (1):\penalty0 129--144, February 2020.
\newblock ISSN 0883-4237.
\newblock \doi{10.1214/19-STS7561}.

\bibitem[Cantwell(2002)]{Cantwell2002-dv}
Brian~J Cantwell.
\newblock \emph{Introduction to symmetry analysis}, volume~1.
\newblock Cambridge University Press, September 2002.
\newblock ISBN 9781139431712.

\bibitem[House(2012)]{House2012-hm}
Thomas House.
\newblock Lie algebra solution of population models based on
  {Time-Inhomogeneous} markov chains.
\newblock \emph{J. Appl. Probab.}, 49\penalty0 (2):\penalty0 472--481, June
  2012.
\newblock ISSN 0021-9002, 1475-6072.
\newblock \doi{10.1239/jap/1339878799}.

\bibitem[Greenman(2022)]{Greenman2022-ok}
Chris~D Greenman.
\newblock Time series path integral expansions for stochastic processes.
\newblock \emph{J. Stat. Phys.}, 187\penalty0 (3):\penalty0 24, April 2022.
\newblock ISSN 0022-4715, 1572-9613.
\newblock \doi{10.1007/s10955-022-02912-8}.

\bibitem[Reis et~al.(2018)Reis, Kromer, and Klipp]{Reis2018-un}
Matthias Reis, Justus~A Kromer, and Edda Klipp.
\newblock General solution of the chemical master equation and modality of
  marginal distributions for hierarchic first-order reaction networks.
\newblock \emph{J. Math. Biol.}, 77\penalty0 (2):\penalty0 377--419, August
  2018.
\newblock ISSN 0303-6812, 1432-1416.
\newblock \doi{10.1007/s00285-018-1205-2}.

\bibitem[Krener(1977)]{Krener1977-ln}
Arthur~J Krener.
\newblock A decomposition theory for differentiable systems.
\newblock \emph{SIAM J. Control Optim.}, 15\penalty0 (5):\penalty0 813--829,
  August 1977.
\newblock ISSN 0363-0129.
\newblock \doi{10.1137/0315052}.

\bibitem[Egri-Nagy(2005)]{Egri-Nagy2005-wu}
Attila Egri-Nagy.
\newblock \emph{Algebraic Hierarchical Decompositions of Finite State Automata
  -- A Computational Approach}.
\newblock PhD thesis, University of Hertfordshire, 2005.

\bibitem[Egri-Nagy et~al.(2008)Egri-Nagy, Nehaniv, Rhodes, and
  Schilstra]{Egri-Nagy2008-ng}
Attila Egri-Nagy, Chrystopher~L Nehaniv, John~L Rhodes, and Maria~J Schilstra.
\newblock Automatic analysis of computation in biochemical reactions.
\newblock \emph{Biosystems.}, 94\penalty0 (1-2):\penalty0 126--134, 2008.
\newblock ISSN 0303-2647, 1872-8324.
\newblock \doi{10.1016/j.biosystems.2008.05.018}.

\bibitem[Note1()]{Note1}
The first reaction is often written as a pseudo-reaction \(\emptyset
  \to B\). However, the resulting mathematical model is equivalent and here we
  prefer the form with a non-zero reactant since using it makes \autoref
  {eq:Zmaster} homogeoneous.

\bibitem[Kendall(1949)]{Kendall1949-kf}
David~G Kendall.
\newblock Stochastic processes and population growth.
\newblock \emph{J. R. Stat. Soc. Series B Stat. Methodol.}, 11\penalty0
  (2):\penalty0 230--282, 1949.
\newblock ISSN 1369-7412.

\bibitem[Kendall(1950)]{Kendall1950-yr}
David~G Kendall.
\newblock An artificial realization of a simple ``{Birth-and-Death}'' process.
\newblock \emph{J. R. Stat. Soc. Series B Stat. Methodol.}, 12\penalty0
  (1):\penalty0 116--119, 1950.
\newblock ISSN 1369-7412.

\bibitem[Lian and Durrett(2018)]{Lian2018-uz}
Tyler Lian and Rick Durrett.
\newblock A new look at multi-stage models of cancer incidence.
\newblock January 2018.

\bibitem[Raj et~al.(2006)Raj, Peskin, Tranchina, Vargas, and Tyagi]{Raj2006-wv}
Arjun Raj, Charles~S Peskin, Daniel Tranchina, Diana~Y Vargas, and Sanjay
  Tyagi.
\newblock Stochastic {mRNA} synthesis in mammalian cells.
\newblock \emph{PLoS Biol.}, 4\penalty0 (10):\penalty0 e309, October 2006.
\newblock ISSN 1544-9173, 1545-7885.
\newblock \doi{10.1371/journal.pbio.0040309}.

\bibitem[Peccoud and Ycart(1995)]{Peccoud1995-se}
J~Peccoud and B~Ycart.
\newblock Markovian modeling of {Gene-Product} synthesis.
\newblock \emph{Theor. Popul. Biol.}, 48\penalty0 (2):\penalty0 222--234,
  October 1995.
\newblock ISSN 0040-5809.
\newblock \doi{10.1006/tpbi.1995.1027}.

\bibitem[Iyer-Biswas et~al.(2009)Iyer-Biswas, Hayot, and
  Jayaprakash]{Iyer-Biswas2009-wh}
Srividya Iyer-Biswas, F~Hayot, and C~Jayaprakash.
\newblock Stochasticity of gene products from transcriptional pulsing.
\newblock \emph{Phys. Rev. E Stat. Nonlin. Soft Matter Phys.}, 79\penalty0 (3
  Pt 1):\penalty0 031911, March 2009.
\newblock ISSN 1539-3755.
\newblock \doi{10.1103/PhysRevE.79.031911}.

\bibitem[Albert(2019)]{Albert2019-xq}
Jaroslav Albert.
\newblock Path integral approach to generating functions for multistep
  post-transcription and post-translation processes and arbitrary initial
  conditions.
\newblock \emph{J. Math. Biol.}, 79\penalty0 (6-7):\penalty0 2211--2236,
  December 2019.
\newblock ISSN 0303-6812, 1432-1416.
\newblock \doi{10.1007/s00285-019-01426-4}.

\bibitem[Vastola et~al.(2021)Vastola, Gorin, Pachter, and
  Holmes]{Vastola2021-hg}
John~J Vastola, Gennady Gorin, Lior Pachter, and William~R Holmes.
\newblock Analytic solution of chemical master equations involving gene
  switching. i: Representation theory and diagrammatic approach to exact
  solution.
\newblock \emph{arXiv [q-bio.SC]}, March 2021.
\newblock \doi{10.48550/arXiv.2103.10992}.

\bibitem[Gilmore(1974)]{Gilmore1974-aq}
Robert Gilmore.
\newblock \emph{Lie Groups, Lie Algebras and Some of Their Applications},
  volume~1.
\newblock Wiley, February 1974.
\newblock ISBN 9780471301790.

\bibitem[Schnoerr et~al.(2016)Schnoerr, Sanguinetti, and
  Grima]{Schnoerr2016-ty}
David Schnoerr, Guido Sanguinetti, and Ramon Grima.
\newblock Approximation and inference methods for stochastic biochemical
  kinetics - a tutorial review.
\newblock August 2016.

\bibitem[Lazarescu et~al.(2019)Lazarescu, Cossetto, Falasco, and
  Esposito]{Lazarescu2019-eo}
Alexandre Lazarescu, Tommaso Cossetto, Gianmaria Falasco, and Massimiliano
  Esposito.
\newblock Large deviations and dynamical phase transitions in stochastic
  chemical networks.
\newblock \emph{J. Chem. Phys.}, 151\penalty0 (6):\penalty0 064117, August
  2019.
\newblock ISSN 0021-9606.
\newblock \doi{10.1063/1.5111110}.

\bibitem[Mendler et~al.(2018)Mendler, Falk, and Drossel]{Mendler2018-ed}
Marc Mendler, Johannes Falk, and Barbara Drossel.
\newblock Analysis of stochastic bifurcations with phase portraits.
\newblock \emph{PLoS One}, 13\penalty0 (4):\penalty0 e0196126, April 2018.
\newblock ISSN 1932-6203.
\newblock \doi{10.1371/journal.pone.0196126}.

\bibitem[Becker et~al.(2020)Becker, Mendler, and Drossel]{Becker2020-uj}
Lara Becker, Marc Mendler, and Barbara Drossel.
\newblock Relation between the convective field and the stationary probability
  distribution of chemical reaction networks.
\newblock \emph{New J. Phys.}, 22\penalty0 (3), 2020.
\newblock ISSN 1367-2630.
\newblock \doi{10.1088/1367-2630/ab73c6}.

\bibitem[Schmid(2012)]{Schmid2012-np}
Rudolf Schmid.
\newblock Infinite dimentional lie groups with applications to mathematical
  physics.
\newblock \emph{Journal of Geometry and Symmetry in Physics}, 1\penalty0 (1),
  2012.
\newblock \doi{10.7546/jgsp-1-2004-54-120}.

\bibitem[Masani(1984)]{Masani1984-ql}
P~R Masani.
\newblock The place of multiplicative integration in modern analysis.
\newblock In \emph{Product Integration with Application to Differential
  Equations}, pages 215--248. Cambridge University Press, December 1984.
\newblock \doi{10.1017/CBO9781107340701.013}.

\bibitem[Dollard and Friedman(1984)]{Dollard1984-to}
John~Day Dollard and Charles~N Friedman.
\newblock \emph{Product Integration with Applications to Differential
  Equations}.
\newblock Cambridge University Press, December 1984.
\newblock ISBN 9781107340701.
\newblock \doi{10.1017/CBO9781107340701}.

\bibitem[Hochstadt(2012)]{Hochstadt2012-xh}
Harry Hochstadt.
\newblock \emph{The Functions of Mathematical Physics}.
\newblock Dover Publications, April 2012.
\newblock ISBN 9780486168784.

\bibitem[Ince(1956)]{Ince1956-yd}
Edward~L Ince.
\newblock \emph{Ordinary Differential Equations}.
\newblock Courier Corporation, January 1956.
\newblock ISBN 9780486603490.

\bibitem[Greenman(2021)]{Greenman2021-qh}
Chris~D Greenman.
\newblock Time series path integral expansions for stochastic processes.
\newblock \emph{arXiv [cond-mat.stat-mech]}, September 2021.
\newblock \doi{10.48550/arXiv.2109.06936}.

\bibitem[Keeling and Ross(2008)]{Keeling2008-hw}
M~J Keeling and J~V Ross.
\newblock On methods for studying stochastic disease dynamics.
\newblock \emph{J. R. Soc. Interface}, 5\penalty0 (19):\penalty0 171--181,
  February 2008.
\newblock ISSN 1742-5689, 1742-5662.
\newblock \doi{10.1098/rsif.2007.1106}.

\bibitem[Al-Mohy and Higham(2010)]{Al-Mohy2010-nf}
Awad~H Al-Mohy and Nicholas~J Higham.
\newblock A new scaling and squaring algorithm for the matrix exponential.
\newblock \emph{SIAM J. Matrix Anal. Appl.}, 31\penalty0 (3):\penalty0
  970--989, January 2010.
\newblock ISSN 0895-4798.
\newblock \doi{10.1137/09074721X}.

\bibitem[Ham et~al.(2020)Ham, Schnoerr, Brackston, and Stumpf]{Ham2020-ij}
Lucy Ham, David Schnoerr, Rowan~D Brackston, and Michael P~H Stumpf.
\newblock Exactly solvable models of stochastic gene expression.
\newblock \emph{J. Chem. Phys.}, 152\penalty0 (14):\penalty0 144106, April
  2020.
\newblock ISSN 0021-9606.
\newblock \doi{10.1063/1.5143540}.

\bibitem[Schnoerr et~al.(2017)Schnoerr, Sanguinetti, and
  Grima]{Schnoerr2017-as}
David Schnoerr, Guido Sanguinetti, and Ramon Grima.
\newblock Approximation and inference methods for stochastic biochemical
  kinetics---a tutorial review.
\newblock \emph{J. Phys. A: Math. Theor.}, 50\penalty0 (9):\penalty0 093001,
  March 2017.
\newblock ISSN 1751-8113.
\newblock \doi{10.1088/1751-8121/aa54d9}.

\bibitem[{\"O}cal et~al.(2019){\"O}cal, Grima, and Sanguinetti]{Ocal2019-ht}
Kaan {\"O}cal, Ramon Grima, and Guido Sanguinetti.
\newblock Parameter estimation for biochemical reaction networks using
  wasserstein distances.
\newblock \emph{J. Phys. A: Math. Theor.}, 53\penalty0 (3):\penalty0 034002,
  December 2019.
\newblock ISSN 1751-8121.
\newblock \doi{10.1088/1751-8121/ab5877}.

\bibitem[Watson et~al.(2020)Watson, Papula, Poon, Wong, Young, Druley, Fisher,
  and Blundell]{Watson2020-fc}
Caroline~J Watson, A~L Papula, Gladys Y~P Poon, Wing~H Wong, Andrew~L Young,
  Todd~E Druley, Daniel~S Fisher, and Jamie~R Blundell.
\newblock {The evolutionary dynamics and fitness landscape of clonal
  hematopoiesis}.
\newblock \emph{Science}, 367\penalty0 (6485):\penalty0 1449--1454, March 2020.
\newblock ISSN 0036-8075.
\newblock \doi{10.1126/science.aay9333}.

\bibitem[Sukys et~al.(2022)Sukys, {\"O}cal, and Grima]{Sukys2022-fg}
Augustinas Sukys, Kaan {\"O}cal, and Ramon Grima.
\newblock Approximating solutions of the chemical master equation using neural
  networks.
\newblock \emph{iScience}, 25\penalty0 (9):\penalty0 105010, September 2022.
\newblock ISSN 2589-0042.
\newblock \doi{10.1016/j.isci.2022.105010}.

\bibitem[Carilli et~al.(2023)Carilli, Gorin, Choi, Chari, and
  Pachter]{Carilli2023-ti}
Maria Carilli, Gennady Gorin, Yongin Choi, Tara Chari, and Lior Pachter.
\newblock Mechanistic modeling with a variational autoencoder for multimodal
  single-cell {RNA} sequencing data.
\newblock January 2023.

\bibitem[nla(2020)]{nlab-Riccati-2023}
Riccati equation.
\newblock \href{https://ncatlab.org/nlab/show/Riccati+equation}{nlab:Riccati equation}, September 2020.
\newblock Accessed: 2023-1-18.

\bibitem[Dattani(2015)]{Dattani2015-kz}
Justine Dattani.
\newblock Exact solutions of master equations for the analysis of gene
  transcription models, 2015.

\bibitem[Dattani and Barahona(2017)]{Dattani2017-uk}
Justine Dattani and Mauricio Barahona.
\newblock Stochastic models of gene transcription with upstream drives: exact
  solution and sample path characterization.
\newblock \emph{J. R. Soc. Interface}, 14\penalty0 (126), January 2017.
\newblock ISSN 1742-5689, 1742-5662.
\newblock \doi{10.1098/rsif.2016.0833}.

\bibitem[Arley and Borchsenius(1944)]{Arley1944-cr}
Niels Arley and Vibeke Borchsenius.
\newblock On the theory of infinite systems of differential equations and their
  application to the theory of stochastic processes and the perturbation theory
  of quantum mechanics.
\newblock \emph{Acta Math.}, 76\penalty0 (3):\penalty0 261--322, September
  1944.
\newblock ISSN 0001-5962, 1871-2509.
\newblock \doi{10.1007/BF02551579}.

\bibitem[Ticciati(1999)]{Ticciati1999-jz}
Robin Ticciati.
\newblock \emph{Quantum Field Theory for Mathematicians}.
\newblock Cambridge University Press, June 1999.
\newblock ISBN 9780521632652.

\end{thebibliography}


\end{document}